\documentclass[prb,twocolumn,superscriptaddress,10pt,longbibliography]{revtex4-1}

\usepackage{graphicx}
\usepackage[caption=false]{subfig}
\usepackage{amsmath,amssymb}
\usepackage[dvipsnames]{color}
\definecolor{myblue}{named}{MidnightBlue}
\usepackage[a4paper=true,colorlinks=true,citecolor=myblue,linkcolor=black,urlcolor=myblue]{hyperref}
\usepackage{placeins}
\usepackage{bbold}
\usepackage{tabularx}
\usepackage{multirow}

\newcommand{\ket}[1]{|#1\rangle}
\newcommand{\bra}[1]{\langle#1|}
\newcommand{\be}{\begin{equation}}
\newcommand{\ee}{\end{equation}}
\newcommand{\ba}{\begin{eqnarray}}
\newcommand{\ea}{\end{eqnarray}}

\def\unity{\mathbb 1}

\begin{document}
\title{\Huge Tunable electromagnetic environment for superconducting quantum bits }

\author{P. J. Jones}
\affiliation{QCD Labs, COMP Centre of Excellence, Department of Applied Physics, Aalto University, P.O. Box 13500, FI-00076 Aalto, Finland}
\author{J. A. M. Huhtam\"aki}
\affiliation{QCD Labs, COMP Centre of Excellence, Department of Applied Physics, Aalto University, P.O. Box 13500, FI-00076 Aalto, Finland}
\author{J. Salmilehto$^*$}
\affiliation{QCD Labs, COMP Centre of Excellence, Department of Applied Physics, Aalto University, P.O. Box 13500, FI-00076 Aalto, Finland}
\author{K. Y. Tan}
\affiliation{QCD Labs, COMP Centre of Excellence, Department of Applied Physics, Aalto University, P.O. Box 13500, FI-00076 Aalto, Finland}
\author{M. M\"{o}tt\"{o}nen}
\affiliation{QCD Labs, COMP Centre of Excellence, Department of Applied Physics, Aalto University, P.O. Box 13500, FI-00076 Aalto, Finland}
\affiliation{Low Temperature Laboratory (OVLL), Aalto University, P.O. Box 13500, FI-00076 Aalto, Finland}

\begin{abstract}
\textbf{We introduce a setup which realises a tunable engineered environment for experiments in circuit quantum electrodynamics. We illustrate this concept with the specific example of a quantum bit, qubit, in a high-quality-factor cavity which is capacitively coupled to another cavity including a resistor. The temperature of the resistor, which acts as the dissipative environment, can be controlled in a well defined manner in order to provide a hot or cold environment for the qubit, as desired. Furthermore, introducing superconducting quantum interference devices (SQUIDs) into the cavity containing the resistor, provides control of the coupling strength between this artificial environment and the qubit. We demonstrate that our scheme allows us to couple strongly to the environment enabling rapid initialization of the system, and by subsequent tuning of the magnetic flux of the SQUIDs we may greatly reduce the resistor-qubit coupling, allowing the qubit to evolve unhindered.}
\end{abstract}

\maketitle

\onecolumngrid

\noindent
*Correspondence and requests for materials should be addressed to J.S. (juha.salmilehto@aalto.fi)

\vspace{0.6cm}

\noindent
The field of circuit quantum electrodynamics\cite{You1,Blais04,Wallraff04,Blais07,You2,You3,Xiang1,Buluta1,Buluta2,Deppe08,Fink08,Liu1,Mariantoni11,Paik11,Niemczyk10} (cQED), wherein quantum mechanical cavity modes couple to non-linear elements such as two-level quantum systems made of electric circuits, has proven to be a very attractive architecture for quantum computing. Indeed, progress has been so impressive that recently fabricated superconducting qubits offer coherence times of such length that prototype fault-tolerant quantum computing is already beginning to look like a genuine possibility\cite{Paik11}. Simultaneously, recent advances in nanoscale engineering have opened up the possibility to enter regimes in which single photons are the dominant heat carriers in mesoscopic systems\cite{Meschke06,Ojanen07,Cleland04, Pascal11,Muhonen12}. In particular, superconductor--insulator--normal-metal (SIN)  junctions provide exquisite temperature control and thermometer sensitivity for small metal islands\cite{Giazotto06}. In Refs.~\onlinecite{CavRevI} and \onlinecite{CavRevII}, these two fields were unified with a study of the photonic heat conduction between two resistors in a superconducting cavity. We present an advancement to these works by demonstrating that bringing together the excellent qubit control of cQED with the sophisticated normal-metal components used in SIN thermometry can offer great benefits for quantum computing. 

Ensuring correct system initialization is vital for efficient operation of quantum algorithms\cite{DiVincenzo00}. For many cQED experiments this initialization can be adequately achieved simply by waiting for a sufficiently large multiple of the qubits natural lifetime, after which it can be assumed to have de-excited with high probability\cite{Wendin06}. As the limits of cQED are pushed harder and harder by increasingly sophisticated techniques, the comparatively slow and imprecise nature of this process is highlighted. Recent works suggest several solutions which offer improvements in performance\cite{Valenzuela06,You4,Grajcar1,Grajcar2,Reed10,Johnson12,Riste12}. In particular, Ref.~\onlinecite{Riste12} has demonstrated initialization by measurement, a promising route which offers a significant speed advantage. In its current realization, however, this approach requires multiple imperfect measurements to rule out the presence of residual cavity excitations. 

Here we report on a simple and possibly more efficient alternative which may also be employed as a complementary technique to further enhance the efficiency of these previously introduced schemes. To this end, we present a setup in which a capacitor is used to divide a superconducting coplanar transmission line into two coupled cavities. There are numerous examples of using such coupled cavities in an advantageous manner in related literature\cite{Zhou1,Zhou2,Zhou3,Liao1}. We place a resistor into one of these cavities, and we introduce a qubit into the second, which retains a high quality-factor. When the frequencies of both the cavities and the qubit are detuned, the cavity housing the qubit is coupled only weakly to the resistor. In this case, both the excited state of the qubit and the photonic excitations of this cavity are well protected from the artificial resistor environment. The inclusion of one or more superconducting quantum interference devices (SQUIDs) into the dissipative cavity, enables us to tune the interaction between the two cavities, and thus modify the probability amplitude of the photonic excitations at the position of the resistor. In this way, we may independently vary the photonic and qubit decay rates over several orders of magnitude. Consequently, it becomes possible to switch the environment of the qubit between a high quality-factor regime, essential for protecting the  qubit during evolution, and a low quality-factor regime, so that the qubit state may be initialized at will. \\ 
\noindent


\noindent
\textbf{\Large Results}\vspace{0.1cm} \\

\begin{figure}
\includegraphics[width=0.5\textwidth]{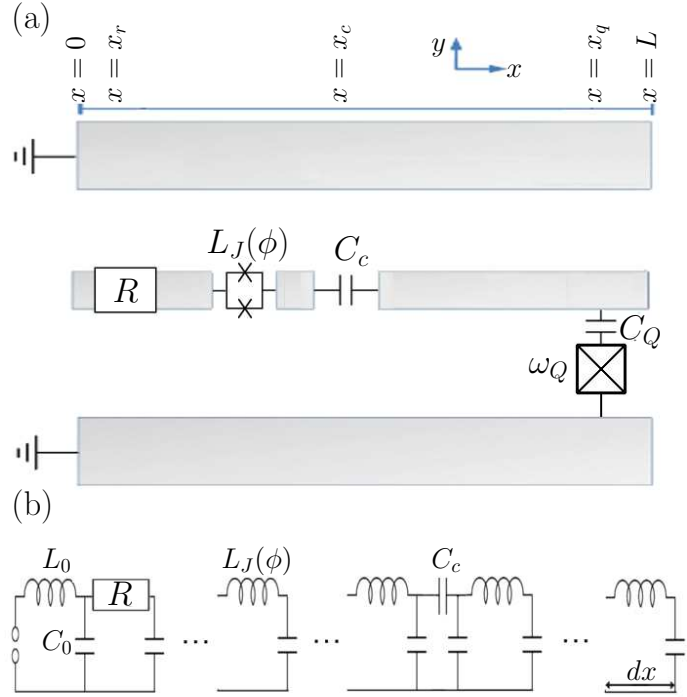}
\caption{\textbf{Schematic and circuit diagrams of the system.} (a) A coplanar waveguide cavity which consists of a center conductor located between two ground planes. The cavity has been modified by adding a resistance $R$ and capacitance $C_c$ at positions $x_r$ and $ x_c$, respectively. In addition, one or more SQUIDs, each with an inductance $L_J(\Phi)$, where $\Phi$ is the magnetic flux penetrating the SQUID loop, are positioned evenly into the left cavity. Our approach makes no assumptions about the type of qubit, however, for the sake of concreteness, we assume a transmon qubit that is capacitively coupled to the resonator with a capacitance $C_Q$ at $x_q$. (b) The cavity may be represented using an infinite number of capacitors ($C_0=c dx$) and inductors ($L_0 = \ell dx)$. The horizontal dots represent many sequences of these $L_0, C_0$ pairs. It is also possible to include series resistors to model the internal losses of the superconductor but these are assumed to be negligible in our setup.} \label{fig:CRQ}
\end{figure}

\noindent
We first revisit work relating to the integration of components within a superconducting coplanar waveguide cavity. Our method follows the usual treatment\cite{Blais04,Yurke84}, but includes a capacitor dividing the cavity into two regions. These regions act as the high- and low-quality cavities depending on their configuration.\\
\vspace{0.1cm} \\
\noindent
\textbf{Modification of Cavity Modes due to a Dividing Capacitor.} We consider a setup in which we insert at position $x_c$, a capacitor of capacitance $C_c$ into the central conducting strip of a high-quality superconducting coplanar waveguide cavity, as shown schematically in Fig.~\ref{fig:CRQ}.
We employ the usual representation of the cavity as a one-dimensional circuit, a model into which the capacitor may be conveniently incorporated. Our starting point to calculate the photonic modes is the classical Lagrangian which may be written in terms of $\Theta(x,t)$, the integral of the charge density in the cavity from $0$ to $x$, as
\begin{equation} \label{eq:FirstLag}
L_{\mathrm{Lagr}}=\int_{0}^{L} \! \left[ \frac{\ell(x)}{2} \left(\frac{\partial \Theta}{\partial t}\right)^{2} -\frac{1}{2c}\left(\frac{\partial \Theta}{\partial x}\right)^{2}-\frac{\Theta^{2}}{2C_c}\delta(x-x_{c})\right] \, dx,
\end{equation}
where the cavity has a length $L$, with a capacitance per unit length $c$, and we allow for a position-dependent inductance per unit length $\ell(x)$ of the form
\begin{equation}
\ell(x)= 
\begin{cases}
\ell_L, &\textrm{$0\le x \le x_{c}$}, \\
\ell_R,  &\textrm{$x_c< x \le L$}.
\end{cases}
\end{equation} 
We look for solutions to the Euler-Lagrange equation of the form $\Theta(x,t)=\sum_{j}X_{j}(x)T_{j}(t)$, where the functions $X_j(x)$ form an orthonormal basis, $\int_0^L \! \ell(x) X_m X_n  \,dx=\delta_{mn}$. The equation takes the form\small
\begin{align} \label{eq:DE}
\frac{1}{c\ell(x)}\frac{1}{X_j(x)}\frac{\partial^{2} X_j(x)}{\partial x^{2}}-\frac{1}{C_c\ell(x)}\delta(x-x_{c}) =  \frac{1}{T_j(t)} \ddot{T_j}(t)=-\omega_j^{2}, 
\end{align} \normalsize
where $\omega_j$ is a constant, i.e., the eigenfrequency of the $j^{\textrm{th}}$ mode. Applying the boundary conditions $X_j(0)=X_j(L)=0$, and continuity at $x_c$ leads to the solution
\begin{align} \label{eq:MP}
&X_j(x)=
\begin{cases}
 \tilde{A}_j\sin[k^{L}_jx], &0\le x \le x_{c}, \\
 \tilde{A}_j \beta_j \sin[k^{R}_j(x-L)], &x_c \le x \le L,
\end{cases}
\end{align}
where $\beta_j=\sin[k_j^{L}x_c]/\sin[k_j^{R}(x_c-L)]$, $k^{L(R)}_j= \omega_j\sqrt{\ell_{L(R)}c}$ and $\tilde{A}_j$ is a normalization constant given by
\begin{align}
\tilde{A}_j=\Biggl \lbrace &\frac{\ell_L}{4k_j^L}\left[2 k_j^Lx_c - \sin(2k_j^L x_c) \right] -\nonumber \\ &\frac{\ell_R \beta_j^2}{4k_j^R}\left[2 k_j^R (x_c-L) - \sin(2k_j^R(x_c-L) \right]\Biggr \rbrace^{-1/2}.
\end{align}
The normalization ensures that $\int \ell(x) [X_j(x)]^{2} dx =1$.

\begin{figure} 
\includegraphics[width=0.5\textwidth]{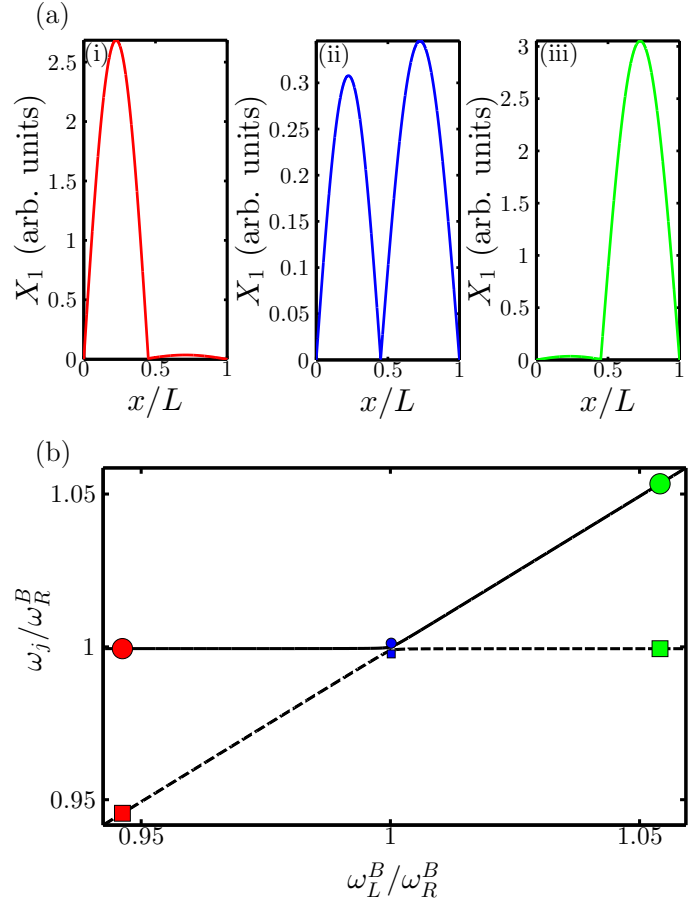}
\caption{\textbf{Mode profiles and frequency spectrum in the divided cavity.} (a) The mode profile of the lowest-energy cavity mode, calculated using Eq.~(\ref{eq:MP}), with a flux through the SQUIDs of (i) $\Phi = 0.35 \Phi_0$, (ii) $\Phi = 0.28\Phi_0$, and (iii) $\Phi = 0.14 \Phi_0$. (b) The frequencies of the two lowest modes, calculated from Eq.~(\ref{eq:wavevec}), as functions of the bare left-cavity frequency to give the spectrum in the frequency range around the left-right cavity resonance. The first and second modes, for the flux values considered in (a), are marked with squares and circles, respectively. For the cavity parameters, we take a cavity of length $L=12 \textrm{ mm}$, and a capacitor positioned at $x_c=0.45L$, with capacitance $C_c=0.5 \textrm{ fF}$. We introduce $n_s$ SQUIDs, each with a critical current $I_{c0} =0.3 \textrm{ }\mu\textrm{A} \times n_s$, into the left side of the cavity. For the center conductor, $c=130 \times 10^{-12}\textrm{ Fm}^{-1}$, and we take the inductance per unit length of the right cavity region to be $\ell_R =3.25 \times 10^{-7}\textrm{ Hm}^{-1}$, resulting in a characteristic impedance $Z_c := \sqrt{\ell_R/c} = 50 \textrm{ }\Omega$. For these parameters, the bare right-cavity frequency has a value $\omega_R^B=2\pi \times 11.66\textrm{ GHz}$, and the interaction frequency is $\Delta\omega^{\textrm{res}}=0.0012 \times \omega^B_R$.} \label{fig:SQRes}
\end{figure}

Integrating Eq.~(\ref{eq:DE}) over an infinitesimal range around $x_c$ provides a condition for the eigenvalues, 
\begin{align}  \label{eq:wavevec}
&\beta_j \omega_j \sqrt{\ell_R c} \cos\left[\omega_j \sqrt{\ell_R c}\left(x_c-L\right)\right]-\nonumber 
\\ &\omega_j \sqrt{\ell_L c} \cos\left(\omega_j \sqrt{\ell_L c} x_c\right)-\frac{c}{C_c}\sin\left(\omega_j \sqrt{\ell_L c} x_c\right) = 0,
\end{align}
which can be utilized to obtain the spectrum $\{\omega_j\}$, for example, numerically. The range of integration is selected for convenience and could be extended over any path that covers $x_c$ to obtain the same spectrum. In the limit $c/C_c \rightarrow \infty$, Eq.~(\ref{eq:wavevec}) gives the solutions $\omega^{B}_L = \frac{1}{\sqrt{\ell_L c}}\frac{j\pi}{x_c}$, and $\omega^B_R = \frac{1}{\sqrt{\ell_R c}}\frac{j\pi}{L-x_c}$ with $j = 0,1,2\dots$ We denote these as the \emph{bare cavity} frequencies, describing uncoupled cavities of length $x_c$, and $L-x_c$, and with inductance per unit length of $\ell_L$, and $\ell_R$, respectively. For finite but still small capacitance $C_c$, placed close to the center of the cavity, the lowest-energy pair of modes have nearly equal energy at the cavity resonance point $\omega_1\approx \omega_2$, separated only by the cavity interaction energy $\hbar \Delta\omega^{\textrm{res}} \ll \hbar \omega^B_R,\hbar \omega^B_L$. Far away from this resonance, the lowest frequencies, $\omega_1 $, and  $\omega_2$, are approximately equal to $\omega^{B}_L$ and $\omega^{B}_R$ such that if $\omega_L^B \ll \omega_R^B$, we find $\omega_1 \approx \omega_L^B$ and $\omega_2 \approx \omega_R^B$, and if $\omega_L^B \gg \omega_R^B$, we obtain $\omega_1 \approx \omega_R^B$ and $\omega_2 \approx \omega_L^B$. This is illustrated in Fig.~\ref{fig:SQRes}(b), which shows the frequencies of the two lowest-energy modes as a function of $\omega_L^B$. The dependence of the spectrum on the cavity parameters is reflected in the mode profiles [Fig.~\ref{fig:SQRes}(a)], in which the amplitude of the fundamental mode is predominantly located in the left-hand side of the cavity for $\omega_L^B \ll \omega_R^B$. In Fig.~\ref{fig:SQRes}(a), the modification of the mode profile is examined with respect to a magnetic flux $\Phi$ threading each of the identical control SQUIDs inserted into the left cavity. For the purpose of interpreting Fig.~\ref{fig:SQRes}(a), it suffices to note that $\ell_L \sim 1/|\cos(\pi\Phi/\Phi_0)|$, where $\Phi_0 = h/2e$ is the flux quantum, [see Eq.~(\ref{eq:aveind})] so that decreasing $\Phi$ corresponds to increasing $\omega_L^B$. Increasing $\omega^{B}_L$ subsequently shifts the mode profile primarily to the right-hand side of the cavity, via an intermediate resonance region in which the amplitude on the left and right is comparable. We therefore refer to these modes as left and right photon excitations, in order to indicate their most probable position in the cavity. This convention is well justified away from cavity-cavity resonance. Close to the resonance point the distinction between left and right is somewhat arbitrary, nevertheless to simplify the discussion we maintain this convention.

We perform the standard quantization of the system, as outlined in the Methods section, to find the decomposition of the operator describing the integral of the charge density, $\hat{\Theta} = \sum_j \hat{\Theta}_j = \sum_j X_j\hat{T}_j$. The quantization yields
\begin{equation} \label{eq:theta}
 \hat{\Theta}_j(x,t) = 
 \begin{cases}
 \vartheta_j  \left[\hat{a}_{j}+\hat{a}^\dag_{j}\right]\sin(k^{L}_j x), &\textrm{$0\le x \le x_{c}$}, \\
 \beta_j \vartheta_j  \left[\hat{a}_{j}+\hat{a}^\dag_{j}\right]\sin[k^{R}_j(x-L)],  &\textrm{$x_c\le x \le L$,}
 \end{cases}
\end{equation}
where $\vartheta_j=\sqrt{\frac{\hbar\tilde{A}_j^{2} }{2\omega_{j}}}$, and $\hat{a}^\dag_j(t)$ and $\hat{a}_j(t)$ are the standard bosonic creation and annihilation operators for the $j^{\textrm{th}}$ photon mode in the Heisenberg picture, respectively. Applying the same notation, we can also write the cavity Hamiltonian as $\hat{H}_{\textrm{cav}}=\sum_j \hbar \omega_j(\hat{a}_j^{\dag} \hat{a}_j +1/2)$. The current operator, $\hat{I}_j=\frac{\partial \hat{\Theta}_j}{\partial t}$, assumes the form 
\begin{equation} \label{eq:Cur}
\hat{I}_j=\begin{cases} i\vartheta_j \omega_j \left[\hat{a}^\dag_{j} -\hat{a}_{j}\right]\sin(k^{L}_j x), &\textrm{$0\le x \le x_{c}$}, \\
i\beta_j\vartheta_j \omega_j \left[\hat{a}^\dag_{j} -\hat{a}_{j}\right]\sin[k^{R}_j (x-L)], & \textrm{$x_c\le x \le L$},
\end{cases}
\end{equation}
and the operator for the voltage in the cavity, $\hat{V}_j=\frac{1}{c}\frac{\partial \hat{\Theta}_j}{\partial x}$, is given by
\begin{align} \label{eq:Vol}
\hat{V}_j= \begin{cases} \frac{k^{L}_j \vartheta_j }{c}  \left[\hat{a}_j+\hat{a}^\dag_j \right]\cos(k^{L}_j x), &\textrm{$0\le x \le x_{c}$}, \\ 
 \frac{k^{R}_j \beta_j \vartheta_j  }{c}  \left[\hat{a}_j+\hat{a}^\dag_j \right] \cos[k^{R}_j (x-L)], &\textrm{$x_c\le x \le L$}.
 \end{cases}
\end{align}
\vspace{0.1cm} \\
\noindent
\textbf{Cavity--Qubit Coupling.} We assume that the qubit is introduced at the position $x_q>x_c$, and consider it to be a Cooper-pair box described by a Hamiltonian $\hat{H}_q = C_{\Sigma} \hat{V}_q^2/2 + E_J \cos \hat{\phi}$ where $C_{\Sigma} = C_J + C_Q + C_0$ is the total capacitance of the Cooper pair box, $C_J$ is the capacitance of the Josephson junction defining the superconducting island, $C_Q$ is the coupling capacitance, $C_0$ is the self-capacitance of the superconducting island, and $\hat{V}_q$ and $\hat{\phi}$ are the operators for the voltage and superconducting phase difference across the junction. The Josephson and charging energies of the box are defined as $E_J$ and $E_C = (2e)^2/C_{\Sigma}$, respectively. The cavity--qubit interaction Hamiltonian is given by $\hat{H}_{c\text{--}q}^{\mathrm{tot}} = -C_Q\sum_j \hat{V}_q\hat{V}_j$ which, after performing the rotating wave approximation,\cite{WallsMilburn} takes the form\cite{Blais04} $\hat{H}_{c\text{--}q}=\sum_j \hbar g_j(\hat{\sigma}^{+}\hat{a}_j+\hat{\sigma}^{-}\hat{a}_j^{\dag}) $. The operators $\hat{\sigma}^{+}$, and $\hat{\sigma}^{-}$ correspond to excitation, and de-excitation of the qubit in the energy eigenbasis, respectively. The effect of the cavity--qubit coupling capacitance on the cavity modes is neglected. By employing Eq.~(\ref{eq:Vol}), we may identify the coupling strength $g_j$ between the qubit and the $j^{\textrm{th}} $ cavity mode as 
\begin{align} \label{eq:coupling}
g_j=- \frac{C_Q}{C_{\Sigma}}\frac{e}{\hbar}\frac{ k^{R}_j \beta_j \vartheta_j}{c}\cos[k^{R}_j (x_q-L)], 
\end{align}
where we assumed that the qubit is in the charging regime $E_C \gg E_J$, and we operate it at the charge degeneracy point. We denote the elementary charge by $e$. In the transmon limit for which $E_J \gg E_C$, the coupling strength becomes\cite{Koch07}
\begin{align} \label{eq:couplingtrans}
g_j=-\sqrt{2}\left(\frac{E_J}{8E_C}\right)^{1/4} \frac{C_Q}{C_{\Sigma}}\frac{e}{\hbar}\frac{ k^{R}_j \beta_j \vartheta_j}{c}\cos[k^{R}_j (x_q-L)]. 
\end{align}
\vspace{0.1cm} \\
\noindent
\textbf{Realizing a Tunable Inductance for the Dissipative Cavity.} Let us make the qubit--cavity interaction tunable by inserting one or more SQUIDs distributed evenly throughout the dissipative cavity. Similar approaches providing an inhomogeneous and tunable coupling between the qubit and the cavity have been previously studied\cite{Ian1,Lopez1}. Moreover, inserting SQUIDs to control the features of the cavity has attracted recent interest\cite{Johansson1,Johansson2,Wilson1}. The setup is shown schematically in Fig.~\ref{fig:CRQ} for the case of a single SQUID. Each SQUID is modelled as a linear inductor $L_{J}(\Phi)=\Phi_{0}/(4 \pi I_{c0}|\cos(\pi \Phi / \Phi_{0} )|)$,  $ I_{c0}$ being the critical current of each identical Josephson junction comprising the SQUID, and $\Phi$ the magnetic flux through the SQUID. We can thus modify the effective inductance of the center conductor of the cavity by varying $\Phi$. In the following, we show that this can have a large impact on the eigenfrequencies and mode profiles, which may in turn affect the cavity--qubit coupling, the eigenstates of the system Hamiltonian, and the transition rates due to the bath coupling substantially. In addition to the setup described in this paper, we expect that tunability can be obtained by inserting a SQUID array at the end of the cavity connecting the center conductor to the ground plane but for simplicity we will refrain from discussing this case.

A simple but effective approximation to account for the effect of the SQUIDs on the system is to consider only the induced change in the average inductance per unit length in the left cavity as
\begin{equation} \label{eq:aveind}
 \ell_L = \frac{1}{x_c}\left[ n_sL_{J}(\Phi)+\ell_L^i x_c                                                                                                                                               \right],
\end{equation}
where $n_s$ is the number of SQUIDs of equal inductance, and $\ell_L^i$ is the intrinsic inductance per unit length of the center conductor. In order to approximate the effect of the SQUIDs as a uniform inductance in this manner, their number must be significantly large but a qualitative correspondence should be achievable even for a small number. In considering only the inductance of the SQUIDs, we have neglected their parallel capacitances $C_{J'}$. This is a valid approximation when the signal frequency is low compared to the plasma frequency of the SQUID, $\omega_p:= 1/\sqrt{L_JC_{J'}}$. In practice, it is possible to reproducibly fabricate SQUIDs with $\omega_p$ higher than $2\pi \times 70 \textrm{ GHz}$; such values are several times higher than the typical frequencies considered here. \\
\vspace{0.1cm} \\
\noindent
\textbf{Interaction of the Resistor with the Cavity Modes.} In this section, we incorporate a resistor of resistance $R$, located at $ x_{r}<x_c$, into our model for the capacitor-modified cavity. The average voltage across the resistor vanishes but the voltage noise from the resistor has the Johnson--Nyquist spectral density 
\begin{equation} \label{eq:specdens}
S_{\delta V}(\omega)= \frac{2 R \hbar \omega}{1-\exp(-\frac{\hbar \omega}{k_{B}T})},
\end{equation}
with $T$ the temperature of the electrons in the resistor. We restrict our analysis only to those systems in which the coupling between the photonic modes and the resistor is weak. This weak coupling criterion is guaranteed provided that the resistor is situated where the current mode amplitude is small and/or the resistor has sufficiently small magnitude in comparison to the characteristic impedance of the cavity $[R\sin^2(k_j^{L}x_r) \ll Z_c:=\sqrt{\ell_R/c}]$. In this limit we can arrive at a relatively simple analytic result for the resistor--cavity interaction Hamiltonian by utilizing the procedure described in Ref.~\onlinecite{CavRevI}, and reviewed in the Methods section, in which the bath couples to the cavity via the voltage fluctuations across the resistor. This results in an interaction term
\begin{align} \label{eq:Hint}
\hat{H}_{\textrm{int}}&=\hat{\Theta}(x_{r})\otimes \delta \hat{V},
\end{align}
where $\delta \hat{V}$ is the operator for the voltage fluctuation over the resistor. 

Applying Fermi's golden rule to an interaction Hamiltonian in the form of Eq.~(\ref{eq:Hint}) yields transition rates between the eigenstates of the cavity--qubit Hamiltonian without the coupling to the environment. The transition rate from the $m$th eigenstate to the $l$th eigenstate is given by
\begin{equation} \label{eq:GTR}
\Gamma_{m\rightarrow l} \approx \frac{|\bra{l} \hat{\Theta}(x_r)\ket{m}|^{2}}{\hbar^{2}} S_{\delta V}(-\omega_{ml}),
\end{equation}
where $\omega_{ml} = (E_l - E_m)/\hbar$, and $ S_{\delta V}(\omega)$ is the spectral density of the voltage noise given in Eq.~(\ref{eq:specdens}). \\
\vspace{0.1cm} \\
\noindent
\textbf{Resistor-Induced Transition Rates.} The total Hamiltonian of the complete cavity--resistor--qubit system, shown in Fig. \ref{fig:CRQ}, is
\begin{equation}
\hat{H}_{\textrm{tot}}=\hat{H}_{\textrm{cav}}+\hat{H}_{\textrm{c--q}}+\hat{H}_{\textrm{q}}+\hat{H}_{\textrm{R}}+\hat{H}_{\textrm{int}}.
\end{equation}
We first consider only the cavity and qubit terms, i.e., the cavity Hamiltonian $\hat{H}_{\textrm{cav}}=\sum_j \hbar \omega_j\left(\hat{a}_j^{\dag} \hat{a}_j +1/2\right)$, the qubit Hamiltonian, for which we assume the usual form, $\hat{H}_{\textrm{q}}=\frac{\hbar \omega_{Q}}{2}\hat{\sigma}_z$, and the cavity--qubit coupling term, $\hat{H}_{\textrm{c--q}}=\hbar \sum_j g_j(\hat{\sigma}^{+}\hat{a}_j+\hat{\sigma}^{-}\hat{a}_j^{\dag})$. Here $\hat{\sigma}_z$ is the Pauli $Z$-operator given in the basis of the two lowest energy eigenstates of the qubit, and acting non-trivially only on the qubit state. In general, the number of modes required for an accurate description of the dynamics depends on the effective bath temperature. Since we assume low temperatures compared with the cavity frequencies, we may consider only the lowest pair of modes in our analysis. We therefore take $\hat{H}_{\textrm{cav}}=\hbar \omega_L(\hat{a}_L^{\dag} \hat{a}_L +1/2)+\hbar \omega_R(\hat{a}_R^{\dag} \hat{a}_R +1/2)$, and $\hat{H}_{\textrm{c--q}}=\hbar g_L(\hat{\sigma}^{+}\hat{a}_L+\hat{\sigma}^{-}\hat{a}_L^{\dag})+\hbar g_R(\hat{\sigma}^{+}\hat{a}_R+\hat{\sigma}^{-}\hat{a}_R^{\dag})$. As above, the subscripts $L$ and $R$ denote the left and right cavity modes, where $\omega_L = \omega_1$, $\omega_R = \omega_2$, if the left cavity is tuned below the cavity resonance point, and $\omega_R = \omega_1$, $\omega_L = \omega_2$, above it. We write the Hamiltonian in the basis $\ket{\sigma,n_L,n_R}$, which is a tensor product of the qubit state $\ket{\sigma}$ ($\sigma = g,e$), and the photon number states, $\ket{n_L}$, and $\ket{n_R}$. Here, $n_L$ is the population of the mode with frequency $\omega_L$, and $n_R$ is the population of the mode with frequency $\omega_R$. Far detuned in the frequency space from either the cavity--cavity, or cavity--qubit resonances, we are able to identify eigenstates of the cavity--qubit system which are, to good approximation, equivalent to the excited states $\ket{g,1,0}$, $\ket{g,0,1}$, and $\ket{e,0,0}$, and hence we refer to these eigenstates as left-photon--like, right-photon--like, and qubit-like, respectively. Close to cavity--qubit resonance, these photon-like eigenvectors will have a significant qubit component and vice-versa, and hence this correspondence is no longer strictly applicable. Nevertheless, studying these eigenvectors still yields much insight into the dynamics of the system.

The resistor acts as an environment for the cavity--qubit system through the interaction term, $\hat{H}_{\textrm{int}}=\hat{\Theta}(x_{r})\otimes \delta \hat{V}$. We take $\hat{H}_{\textrm{int}}$ to be a weak perturbation which stimulates transitions between the different eigenstates of the pure system without the environment due to the presence of the resistor. The system is linear in the weak coupling limit and hence there is no direct interaction between the modes. The interaction Hamiltonian therefore remains in the form $\hat{H}_{\textrm{int}}=\hat{\Theta}(x_{r})\otimes \delta \hat{V}$, where $\hat{\Theta}(x_{r})= X_L(x_r)\hat{T}_L+X_R(x_r)\hat{T}_R$ has two components in Eq.~(\ref{eq:Hint}), one for each mode. We  access the solution of this system numerically, and calculate the transition rates between each of the eigenvectors using Eq.~(\ref{eq:GTR}).

\begin{figure} 
\includegraphics[width=0.5\textwidth]{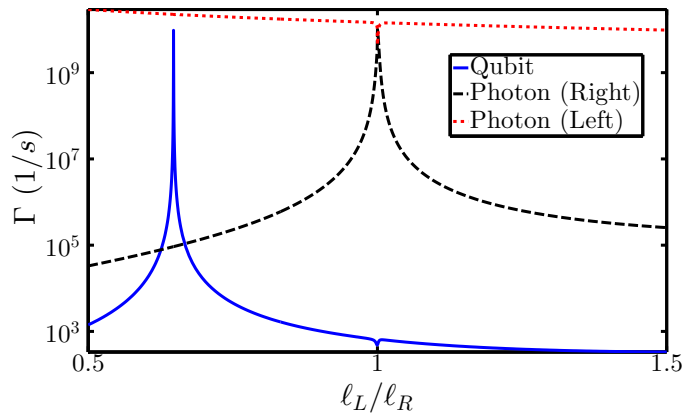}
\caption{\textbf{Decay rates vs. inductance per unit length of the left cavity.} Decay rates to the ground state, for a system prepared in an initial eigenstate of the cavity-qubit system corresponding, away from resonance, to either a qubit excitation (solid line), the excitation of a photon in the right cavity (dashed line), or in the left-cavity (dotted line), as functions of the average inductance per unit length of the left cavity, $\ell_L$. We observe a peak in the decay rate of the right-photon--like state at cavity--cavity resonance, and a peak at the left-cavity--qubit resonance for the decay rate of the qubit-like state. The resistor has an effective temperature $T= 10\textrm{ mK}$, and a resistance of $R=230\textrm{ }\Omega$. The angular frequency of the qubit is held fixed at $2\pi \times 15.91\textrm{ GHz}$. The cavity length is $L=12 \textrm{ mm}$, and includes a capacitor positioned at $x_c=L/2$, with capacitance $C_c=0.5 \textrm{ fF}$, and cavity interaction energy $\hbar\Delta\omega^\textrm{res}=2\pi\hbar\times 16.73$~MHz.  For the center conductor, we have $c=130 \times 10^{-12}\textrm{ Fm}^{-1}$, with a characteristic impedance of the right cavity $Z_c := \sqrt{\ell_R/c} = 50 \textrm{ }\Omega$, resulting in a bare right-cavity frequency $\omega_R^B=2\pi \times 12.82\textrm{ GHz}$.} \label{fig:ellvsT}
\end{figure}

Figure~\ref{fig:ellvsT} shows the transition rates for the qubit-like and photon-like eigenstates as functions of $\ell_L$. 
The behaviour of the photon-like transition rates can be explained by referring to Fig.~\ref{fig:SQRes}(a). If the cavities are far detuned from cavity--cavity resonance, the eigenstates of $\hat{H}_{\mathrm{cav}}$ correspond closely to a photonic excitation in either the left or the right of the dividing capacitor, $C_c$, as evident by the mode amplitudes being predominantly on either side. The left-photon--like eigenstates always have a large $\ket{g,1,0}$ component, and thus high amplitude at the position of the resistor. These left-photon--like states therefore decay very quickly. In contrast, for most values of $\ell_L$, the right-photon--like states are approximately equivalent to the states $\ket{g,0,1}$, which have a much smaller amplitude at $x_r$, and thus the right--photon--like states are protected from decay. As the cavity--cavity resonance point is approached, the right-photon--like eigenstate obtains an increased amplitude at the resistor and, hence decays rapidly, resulting in a peak of several orders of magnitude in the decay rate. 

\begin{figure} 
\includegraphics[width=0.5\textwidth]{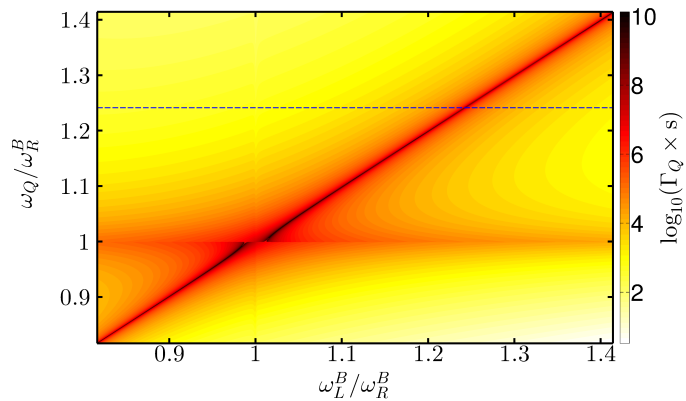}
\caption{\textbf{Contour plot of the qubit--like decay rate.} Decay rate to the ground state, for a system prepared in an initial eigenstate of the cavity-qubit system corresponding, away from resonance, to a qubit excitation, as a function of the frequencies of both the qubit, $\omega_Q$, and bare left cavity, $\omega_L^B$. We plot the logarithm of the decay rates, with darker areas representing higher rates. The cavity parameters are otherwise identical to those used in Fig.~\ref{fig:ellvsT}. The horizontal dashed line marks the trace covered in Fig.~\ref{fig:ellvsT}.}  \label{fig:ellvsQ}
\end{figure}

The qubit-like state couples to the resistor only indirectly via the cavities. To achieve a strong qubit--photon coupling requires a large mode amplitude at the position of the qubit in the right cavity, while to achieve a large photon--resistor coupling requires a large mode amplitude at the position of the resistor in the left cavity. In most frequency regimes, it is therefore not possible to simultaneously strongly couple the qubit to the mode, and the mode to the resistor. Only if the qubit-like eigenstate  is in resonance with the left cavity, and hence a superposition with a significant $\ket{g,1,0}$ component, does this eigenstate decay quickly. We would like to emphasize that such decay is achieved at resonance even if the cavity-cavity detuning is negative. In Fig.~\ref{fig:ellvsQ}, the decay rate for the qubit-like eigenstate is shown as a function of both $\omega_Q$, and $\omega_L^B$. The behaviour of the qubit rate shown in Fig.~\ref{fig:ellvsT}, can be recovered by taking a horizontal trace through Fig.~\ref{fig:ellvsQ}, and is marked with a dashed line. Furthermore, the additional degree of freedom provided by selecting the qubit frequency, equivalent to scanning vertically through Fig.~\ref{fig:ellvsQ}, gives access to an even wider range of decay rates. Most notably, around full resonance, $\omega_Q = \omega_L^B=\omega_R^B$, qubit lifetimes of only a few nanoseconds are predicted. In summary, we can move the system to the left-cavity--qubit resonance point and quickly reset the state of the qubit, or to the cavity--cavity resonance point in order to reset the photon state. By utilizing also the tunability of the qubit, we may bring both cavities and the qubit into joint resonance, and rapidly initialise the entire system. Alternatively, by detuning both the cavities from each other, as well as the qubit, we can simultaneously protect both the photon in the right cavity, and the qubit from decay.

\begin{figure} 
\includegraphics[width=0.5\textwidth]{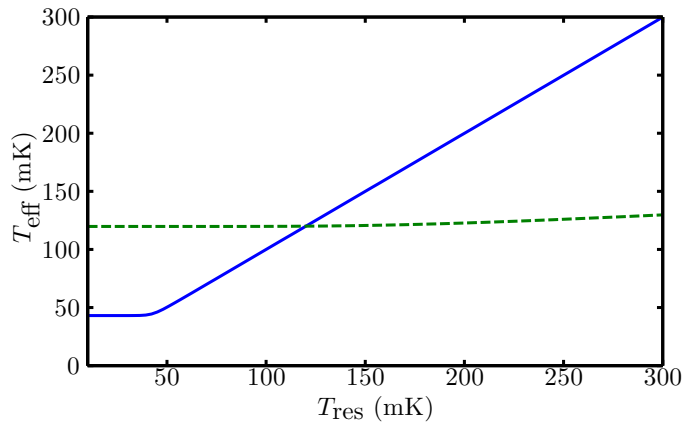}
\caption{\textbf{Effective qubit temperature vs. resistor temperature.} Effective temperature of the qubit as a function of the temperature of the resistor, which may be manipulated using SIN tunnel junction thermometry. The results are shown for bare left-cavity frequencies of $\omega_L^B=1.002 \times \omega_Q$ (solid line) and $\omega_L^B=0.75 \times \omega_Q$ (dashed line) resulting in zero-temperature relaxation rates of $9.59 \times 10^9$ $\mathrm{s}^{-1}$ and $1.01 \times 10^5$ $\mathrm{s}^{-1}$, respectively. The frequency $\omega_L^B=1.002 \times \omega_Q$ is selected to yield the maximum decay rate for the qubit--like eigenstate. In calculating the total excitation and de-excitation rates, we include an intrinsic qubit relaxation rate of $\Gamma_-^i=1.0\times 10^5\textrm{ s}^{-1}$, and excitation rate of $\Gamma_+^i=1.7\times 10^2\textrm{ s}^{-1}$, to give an intrinsic qubit temperature of $T^i=120\textrm{ mK}$. The parameters are otherwise identical to those used in Fig.~\ref{fig:ellvsT}.} \label{fig:Teff}
\end{figure}

In addition to tuning the frequency of the qubit and the left cavity, our setup can utilize SIN thermometry, which gives us control of the resistor temperature, $T$. In the simplest application of such thermometry, the resistor is implemented using a voltage biased SIN junction\cite{Muhonen12,Giazotto06} where the normal metal island acting as the resistor is formed at the center conductor. The existence of the gap in the density of states of the superconductor prevents electrons in this energy range from tunnelling out of the normal metal. By applying a suitable bias voltage, we can effectively modify the Fermi distribution of the island allowing for the temperature control. In general, there is back-action from the SIN junction to the cavity but this can be made insignificant with current technologies\cite{Muhonen12,Giazotto06}. The effect of the resistor temperature on the qubit decay can be studied if we define an effective qubit temperature~\cite{Clerk10}
\begin{equation}
T_{\textrm{eff}} = \frac{\hbar \omega_Q}{k_B \ln\left(\frac{\Gamma_-}{\Gamma_+}\right)},
\end{equation}
where we assume that the system is in quasi-equilibrium. The rates $\Gamma_{\pm}$ refer to the excitation and relaxation rates for the qubit, respectively. In calculating $T_{\textrm{eff}}$, we may phenomenologically account for external noise sources by adding intrinsic excitation and relaxation rates $\Gamma_{\pm}^i$ to the rates calculated with Eq.~(\ref{eq:GTR}) which result from the presence of the resistor. In Fig.~\ref{fig:Teff}, we show the effective qubit temperature as a function of the resistor temperature for two cases of a bare left-cavity frequency close to resonance with the qubit, or detuned from the qubit frequency. We note that by changing the resistor temperature, the effective qubit temperature is almost constant in the detuned case, as the qubit is effectively decoupled from the artificial environment. If the left cavity is brought close to resonance with the qubit, where the artificial environment couples strongly to the qubit, $T_{\textrm{eff}}$ may be decreased drastically.
 \\
\vspace{0.2cm} \\
\noindent 
\textbf{\Large Discussion} \\
\vspace{0.1cm} \\
\noindent
In summary, we have presented a system in the framework of circuit quantum electrodynamics for which an engineered resistor is able to act as the dominant environment for a qubit. Unlike the natural environment, which can have unpredictable effects and may also be a source of particularly problematic high-frequency noise, we can conveniently control this artificial environment. We demonstrate that both the photon and qubit lifetimes can separately be tuned over many orders of magnitude. This allows for rapid initialization, or essentially unhindered operation of the cavity--qubit system as desired. A similar operating principle was also recently studied for a system utilizing coupled LC resonators~\cite{Jones4} complementing our results. By employing techniques from SIN tunnel junction thermometry, we may not only determine the temperature of the artificial environment with high accuracy, but are also able to control it. With these techniques, variations of the resistor temperature over a range of several hundred millikelvins are attainable~\cite{Giazotto06,Muhonen12}, providing control over the effective temperature of the qubit. 

Our work presents an efficient means to utilize normal-metal components within a cQED framework, and is likely to inspire further developments in this direction in order to expand the toolbox of cQED. Though undoubtedly experimentally challenging, the parameters used in our analysis have been selected in line with what is currently achievable\cite{Goppl08}, and thus we expect these results to be experimentally reproducible.\vspace{0.2cm}

\noindent
\textbf{\Large Methods}
\vspace{0.1cm} \\
\noindent
\textbf{Quantization of the Cavity.} We begin by rewriting the Lagrangian of Eq.~(\ref{eq:FirstLag}) in the form
\begin{equation} \label{eq:Lag}
L_{\mathrm{Lagr}}=\sum_{j}\left[\frac{1}{2} \left(\frac{\partial T_{j}(t)}{\partial t}\right)^{2} -\frac{\omega_j^{2}}{2}T_{j}(t)^{2}\right],
\end{equation}
where Eq.~(\ref{eq:DE}) and the orthonormality of the modes have been used. The Hamiltonian is then constructed as the Legendre transform of the Lagrangian density, that is $H=\sum_{j}\left[\dot{T}_{j}^{2}/2+\omega_j^{2}T_{j}^{2}/2\right]$. 
Promoting $\dot{T}_{k}$ and $T_{j}$ to operators with the requirement $\left[\hat{T}_{j},\dot{\hat{T}}_{k} \right]=i\hbar \delta_{j,k}$,
results in a Hamiltonian which is diagonal in the cavity number basis, $\hat{H}=\sum_j \hbar \omega_j[\hat{a}^\dag_j(t)\hat{a}_j(t)+1/2]$,
where we defined 
\begin{align}
\hat{T}_{j}=\sqrt{\frac{\hbar }{2\omega_{j}}} \left[\hat{a}_j(t)+\hat{a}^\dag_j(t) \right], \label{eq:cp}\\
\dot{\hat{T}}_{j}=i\sqrt{\frac{\hbar \omega_{j}}{2}}\left[\hat{a}^\dag_j(t)-\hat{a}_j(t) \right], \label{eq:cm} 
\end{align}
where $\hat{a}^\dag_j(t)$ and $\hat{a}_j(t)$ are the bosonic creation and annihilation operators of the $j^{\textrm{th}}$ photon mode. \\
\vspace{0.1cm} \\
\noindent
\textbf{Resistor--Cavity Coupling.} We calculate the interaction Hamiltonian as the difference in the energy of the resistor--cavity system for the instances in which the resistor does and does not interact with the modified cavity. 
The  Hamiltonian can be found by integrating the energy density in the cavity along its length. In the non-interacting situation, this process results in the Hamiltonian  $\hat{H}_{0} = \hat{H}_{\textrm{cav}}+\hat{H}_{R} =\int^{L}_{0} \! ({c\hat{V}^{2}}+{\ell \hat{I}^{2}})/2  \, dx+\hat{H}_{R}  $, where $\hat{I}$ is the current operator, and $\hat{V}$ the voltage operator in the cavity, given by Eqs.~(\ref{eq:Cur}) and (\ref{eq:Vol}), respectively. The resistor Hamiltonian is $\hat{H}_{R}$, which we may consider to describe an infinite bath of harmonic oscillators, though the explicit form of $\hat{H}_R$ will play no role in our analysis. For the case in which the resistor and the cavity interact, the calculation is marginally more involved. We proceed similarly to the non-interacting setup and write
\begin{align}
\hat{H}(t) =& \hat{H}_{R} + \frac{c}{2}\int^{x_r}_{0} \! ( \hat{V}_{\textrm{cav}}\otimes \unity_{\textrm{res}}+\unity_{\textrm{cav}} \otimes \delta \hat{V}_{L})^{2} \, dx \nonumber \\  
&+ \frac{c}{2}\int^{L}_{x_r} \! (\hat{V}_{\textrm{cav}}\otimes \unity_{\textrm{res}}+\unity_{\textrm{cav}} \otimes \delta \hat{V}_{R})^{2} \, dx \nonumber \\  &+ \frac{\ell}{2}\int^{L}_{0} \! \hat{I}^2_{\textrm{cav}} \, dx.
\end{align}
Here $ \delta \hat{V}_{L(R)} $ is the time-dependent resistor-induced shift in the voltage on the left (right) side of the resistor. As in Ref.~\onlinecite{CavRevI}, the second order terms in $\delta \hat{V}_{L(R)}$ as well as the fluctuations in the current operator, $\delta I  \ll \delta V/R$ are neglected due to the fact that the the fluctuations are weak and the resistor is situated close to the end of the cavity. Denoting the voltage operator on the left of the capacitor by $\hat{V}^{\textrm{cav}}_{L}$ and on the right by $\hat{V}^{\textrm{cav}}_{R}$,
\begin{align}
\hat{H}_{\textrm{int}}&=\hat{H}(t)-\hat{H}_{0}\nonumber \\&=c\Bigl[\int_{0}^{x_{r}} \! \hat{V}^{\textrm{cav}}_{L} \otimes \delta \hat{V}_{L} \, dx + \int_{x_{r}}^{x_{c}} \! \hat{V}^{\textrm{cav}}_{L} \otimes \delta \hat{V}_{R} \, dx \nonumber \\
 &+ \int_{x_{c}}^{L} \! \hat{V}^{\textrm{cav}}_{R} \otimes \delta \hat{V}_{R}\, dx \Bigr].
\end{align}
Performing the integrals we obtain Eq.~(\ref{eq:Hint}), the weak-coupling interaction Hamiltonian
\begin{align} 
\hat{H}_{\textrm{int}}&=\hat{\Theta}(x_{r})\otimes \delta \hat{V},
\end{align}
where we have defined the total voltage fluctuation $\delta \hat{V}\equiv \delta \hat{V}_{L}-\delta \hat{V}_{R}$ and used the continuity of $\hat{\Theta}$ at $x_{c}$.
We note, by comparison with Ref.~\onlinecite{CavRevI}, that the introduction of the capacitor does not affect the functional form of the interaction Hamiltonian although it affects $\hat{\Theta}$.

\vspace{0.6cm}

\noindent
\textbf{Acknowledgements}\\
\noindent
We have received funding from the Academy of Finland through its Centres of Excellence Program under Grant No. 251748, in addition to Grants No. 138903, and No. 135794. The V\"ais\"al\"a Foundation, the Finnish National Doctoral Programme in Materials Physics (NGSMP), and the European Research Council under Grant Agreement No. 278117 (SINGLEOUT) are acknowledged for financial support. We acknowledge CSC – IT Center for Science Ltd. for the allocation of computational resources. \\

\noindent
\textbf{Author Contributions}\\
\noindent
P.J.J. and J.S. wrote the manuscript text. P.J.J. performed the numerical calculations. J.A.M.H. and K.Y.T. provided important ideas at various stages throughout the project. M.M. provided extensive feedback, and corrections for the manuscript. All authors discussed the results. \\

\noindent
\textbf{Additional Information}\\
\noindent
Competing financial interests: The authors declare no competing financial interests.

\end{document}